\documentclass[11pt]{article}
\usepackage{geometry}                
\usepackage{graphicx}
\usepackage{amssymb}
\usepackage{epstopdf}
\title{The family of Quasi-satellite periodic orbits in the circular co-planar RTBP}

\author{Alexandre Pousse\footnote{IMCCE, Observatoire de Paris, UPMC Univ. Paris 06, Univ. Lille 1, CNRS, 77~Av. Denfert-Rochereau, 75014 Paris, France }, \, Philippe Robutel$^*$,\, Alain Vienne$^*$}

\date{September 15, 2014}

\newcommand\lam{{\lambda} }

\newcommand\Gam{{\Gamma} }
\newcommand\eps{{\varepsilon} }

\newcommand\cF{{\cal F} }

\newcommand{\be}{\begin{equation}}
\newcommand{\ee}{\end{equation}}

\begin{document}

\maketitle

\begin{abstract}
In the circular case of the coplanar Restricted Three-body Problem, we studied how the family of quasi-satellite (QS) periodic orbits allows to define an associated libration
center.
Using the averaged problem, we highlighted a validity limit of this one: for QS orbits with low eccentricities, the averaged problem does not correspond to the real problem. 
We do the same procedure to $L_3$, $L_4$ and $L_5$ emerging periodic orbits families and remarked that for very high eccentricities $\cF_{L_4}$ and $\cF_{L_5}$ merge with $\cF_{L_3}$  which bifurcates to a stable family.
\end{abstract}

\section{Introduction}
In the framework of the Restricted Three-body Problem (RTBP), we consider a primary whose mass is equal to one, a secondary in circular motion with a mass $\eps$ and a massless third body; the three bodies are in coplanar motion and in co-orbital resonance configuration.  
We actually know three classes of regular co-orbital motions: in rotating frame with the planet, 
the tadpoles orbits (TP) librate around Lagrangian equilibria $L_4$ or $L_5$; 
the horseshoe orbits (HS) encompass the three equilibrium points $L_3$, $L_4$ and $L_5$;
the quasi-satellites orbits (QS) are remote retrograde satellite around the secondary, but outside of its Hill sphere .

Contrarily to TP orbits, the QS orbits do not emerge from a fixed point in rotating frame as these orbits are always eccentric. 
Thus, we can reformulate QS definition in terms of elliptical heliocentric orbits which librate, in rotating frame, around the planet position. The QS libration center (l.c.) can be materialized in rotating frame by a one-parameter family of periodic orbits.
Our goal is to study this family of periodic orbits in the plane.

\begin{figure}[!ht]
\begin{center}
 \includegraphics[ width=15cm]{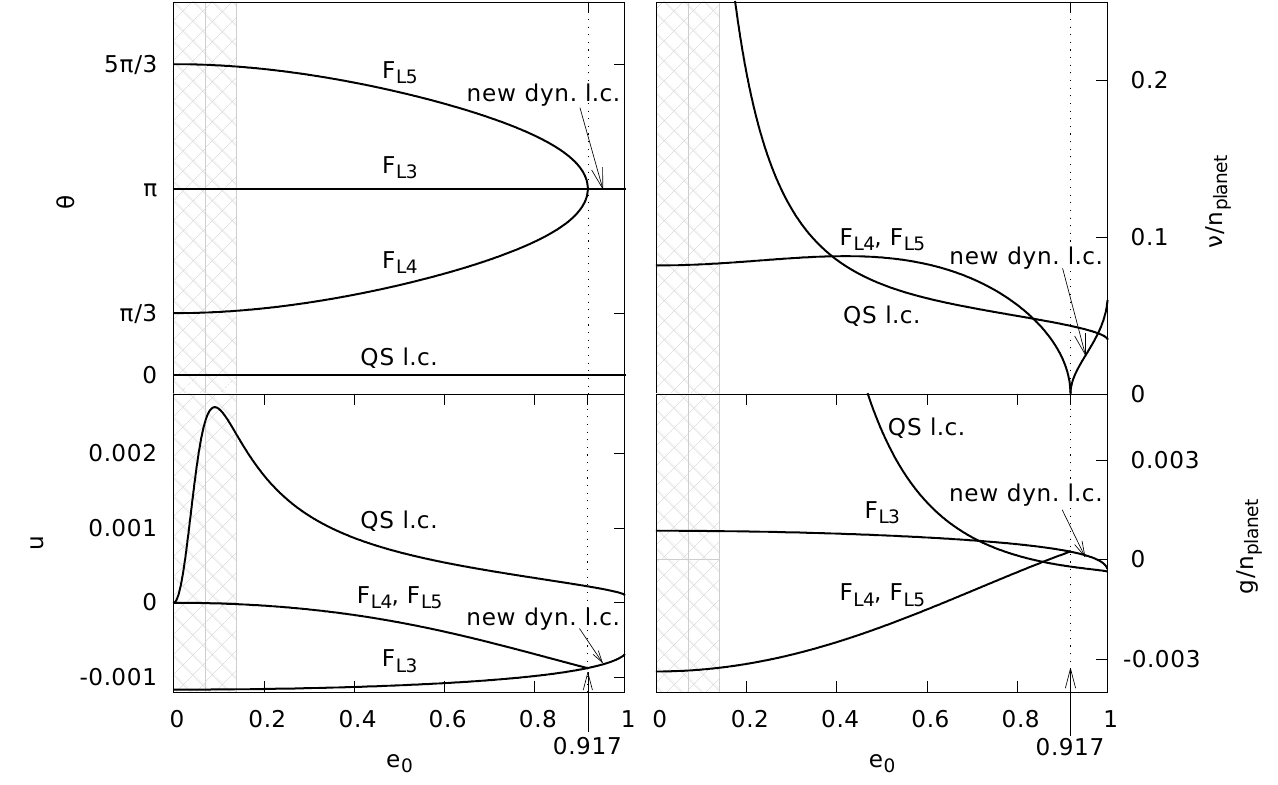}
 \caption{Fixed points and frequencies evolution versus $e_0$ for $\eps = 0.001$ (Sun-Jupiter system) in the averaged coplanar RTBP in circular case.}
   \label{fig1}
\end{center}
\end{figure}

\section{Method}
With respect to the primary, we denote $(a',\lam')$ the elliptic elements of the secondary and  $(a, e, \omega, \lam)$  those of the third body.
To reduce the dimension of the problem to $4$, we introduce the resonant angle $\theta = \lam - \lam'$ and average the Hamiltonian with respect to $\lam'$.
We use the method developed by \cite{NeThFeMo02} which is valid for all values of the eccentricity, to compute the averaged Hamiltonian and equations of motions.

In addition to $\theta$ and $\omega$, we use the variables $u = (\sqrt{a} - \sqrt{a'})/\sqrt{a'}$ and $\Gam = \sqrt{a}(1-\sqrt{1-e^2})$ which is a first integral.
The variable $\omega$ being ignorable, the problem possesses one degree of freedom -- $(\theta, u)$ -- for a given $\Gamma$.
Instead of using $\Gam$, it is convenient to introduce $e_0$, the eccentricity value when $u = 0$. 

In these coordinates, the QS family is represented by a family of fixed points parametrized by $e_0$. When it is stable, each of these equilibria, processes a eigenfrequency $\nu$ that corresponds to the libration frequency around this point in the plan $(\theta,u)$, and a second one $g$ in the normal direction associated to $(\omega, \Gamma)$.  This last frequency is also the precession rate of the third body's pericenter. 

We developed a numerical method to find this fixed point near the origin of the $(\theta, u)$ plane in each $e_0$ and calculated the two frequencies on it.
To compare QS with TP and HS, we do the same procedure to $\cF_{L_3}$, $\cF_{L_4}$ and $\cF_{L_5}$ (the family of periodic orbits that originate from $L_3$, $L_4$ and $L_5$).
The results obtained are presented in Fig. \ref{fig1}.


\section{Discussion}
Since QS orbits with low values of eccentricity  imply close encounters with the secondary, the averaged problem does not represent the real problem in these conditions   (phenomena also observed in \cite{RoPo13} in the planetary Three-body problem).  
In Fig. \ref{fig1}, we remark that when $e_0< 0.18$, the frequencies of the QS family are of the same order than the mean motion. This gives us a region where QS motion can not being studied with the averaged problem (hatched domain). 
We also present a particular orbit for $e_0$ close to $0.8352$ : $g$ crosses zero and highlights a QS frozen ellipse in the fixed frame.


For a very high eccentricity ($e_0 >0.917$), an unexpected result is that $\cF_{L_4}$ and $\cF_{L_5}$ merge with $\cF_{L_3}$  which bifurcates to a stable periodic orbits family. 
This implies the disappearance of TP and HS and the appearance of a new type of orbits.
These orbits librate around the point diametrically opposed to the secondary, relative to the primary.

\end{document}